# NMR study of the magnetic and metal-insulator transitions in $Na_{0.5}CoO_2$: a nesting scenario

Julien Bobroff,[1] Guillaume Lang,[1] Henri Alloul,[1] Nicole Blanchard,[1] Gaston Collin[2]

[1]Laboratoire de Physique des Solides, UMR8502, Université Paris XI, 91405 Orsay, France
[2]LLB, CE-Saclay, CEA-CNRS, 91191 Gif Sur Yvette, France

Co and Na NMR are used to probe the local susceptibility and charge state of the two Co sites of the Na-ordered orthorhombic $Na_{0.5}CoO_2$. Above $T_N = 86K$, both sites display a similar T-dependence of the spin shift, suggesting that there is no charge segregation into $Co^{3+}$ and $Co^{4+}$ sites. Below $T_N$, the magnetic long range commensurate order found is only slightly affected by the metal-insulator transition (MIT) at $T_{MIT} = 51K$. Furthermore, the electric field gradient at the Co site does not change at these transitions, indicating the absence of charge ordering. All these observations can be explained by two successive SDW induced by nestings of the Fermi Surface specific to the x=0.5 Na-ordering.

The $Na_xCoO_2$ cobaltates consist of two-dimensional Co triangular layers of which doping can be varied by changing the Na content. Even though doped layers of transition ions seems similar to the $CuO_2$ planes of High-$T_C$ cuprates, their physical properties are strikingly different. The $Na_1CoO_2$ composition is a band-insulator made of filled-shell non magnetic $Co^{3+}$ [1]. At *x<1*, the system becomes metallic, but only specific Na compositions can be obtained, which correspond to some Na orderings as observed by crystallography. Such Na orderings may turn into charge orderings among Co planes as the Co valence state depends on the Co position relative to the Na ion [2,3]. For *x>0.6*, strong electronic correlations are observed between Co, revealed by a Curie-Weiss *1/(T+Θ)* susceptibility with *Θ~100*K. At lower Na content i.e. higher doping, the compound becomes more metallic as evidenced by a Pauli-like susceptibility and an increase of the Drude weight in optics, except for *x=0.5* [4]. At *x=0.3*, the system can even be superconducting when water molecules are intercalated between Na and $CoO_2$ layers [5]. However, this apparent smooth evolution in the phase diagram from a correlated to a non-correlated metal shows a sharp discontinuity at *x=0.5*. At this composition, Na orders in an orthorhombic superstructure commensurate with the Co lattice. $Na_{0.5}CoO_2$ is a poor metal, with resistivity about twenty times larger than at other Na contents, and shows a metal-insulator transition at $T_{MIT}=51$K [2]. In addition, it displays a long range magnetic order below $T_N=86$K, a Neel temperature much higher than those measured at higher doping contents [2,6]. At the same $T_N$, the Hall coefficient and thermoelectric power change sign, while a small kink is observed on resistivity [2]. Such peculiarities have been interpreted in terms of a charge ordering (CO) in the Co layers induced by the Na orthorhombic order [2,4,7]. Assuming Co ion is a filled-shell *3+* state when it locates at the vertical of a Na ion, and a *4+ S=1/2* state otherwise, the Na superstructure should result in a charge segregation between rows of non magnetic $Co^{3+}$ and magnetic $Co^{4+}$, leading to the observed insulating magnetic behaviour. However, this interpretation fails to explain why CO takes place only at $T_{MIT}$, while Na order is already observed at room T. It is also hard to reconcile with the fact that magnetic order takes place at $T_N \neq T_{MIT}$.

In order to state on the actual existence of CO, we have performed a $^{59}$Co and $^{23}$Na NMR study which allows to differentiate the two Co sites, and to give indications on their valence state through their paramagnetic and quadrupolar parameters. Our results allow us to establish that the simple ionic charge segregation picture into $Co^{3+}$-$Co^{4+}$ does not occur neither at $T_N$, nor at $T_{MIT}$. Instead, we propose an interpretation in terms of a metallic band in which two successive Spin Density Wave transitions are driven by two nestings of the Fermi Surface. This scenario reconciles transport and magnetic measurements.

NMR measurements were carried out on orthorhombic $Na_{0.5}CoO_2$ obtained similarly to that of Ref. [6]. Crystallites were aligned along their c axis in epoxy in a *7* T applied field. NMR spectra were obtained by either sweeping the field or the frequency and reconstructing the spectrum through Fourier Transform recombinations. The nuclear resonance is a probe of the local spin susceptibility through the shift $K_c = A_{hf}^c \chi / \mu_B + K_c^{orb}$ where $\chi$ is the electronic susceptibility, $A_{hf}^c$ is the hyperfine coupling and $K_c^{orb}$ is an orbital T-independent term. It is also sensitive to the surrounding charges through the electric field gradient (EFG) proportional to the quadrupolar frequencies $v_Q^{j=a,b,c} \propto Q \partial^2 V / \partial j^2$ where Q is the quadrupolar moment. For a nuclear spin *I*, these parameters are measured through the detection of the different resonant frequencies for each transition $m \leftrightarrow m-1$ (m=I, I-1, ..., -I+1), following:

$$v_{m \leftrightarrow m-1}^c = v_0(1+K_c) + (0.5-m)v_Q^c + a(v_Q^a - v_Q^b)^2 / v_0$$

where the field $H_0$ is applied along *c* axis, with $v_0 = \gamma/2\pi H_0$, $\gamma$ is the gyromagnetic ratio, *a* depends on *m* and *I*.

Since the orbital shift is almost negligible in the case of $^{23}$Na NMR, the Na shift is directly proportional to the average Co layer spin susceptibility through hyperfine couplings between Na and Co [3]. Fig.1 displays this shift as determined from the frequency of the $-\frac{1}{2} \leftrightarrow \frac{1}{2}$ transition corrected from second order quadrupolar effects. For *T>210* K, the line shape is narrow and no quadrupolar satellite is detected. This results from the occurrence of Na ionic diffusion, as also revealed by a sharp minimum in transverse relaxation time $T_2$ observed as well in $Na_{0.7}CoO_2$ [8]. At *T<210*K, the motion of the Na ions freezes, and the two Na crystallographic sites can be resolved, but are found to display similar shifts. We shall go back to the low T magnetic regime later. The macroscopic susceptibility $\chi_m$ is plotted on Fig.1 as well, using a hyperfine coupling scaling factor close to that of $Na_{0.7}CoO_2$. The Na shift shows the same T-dependence as $\chi_m$ as expected. This allows to extract the orbital contribution $\chi_{orb}= 2.1\ 10^{-4}$ emu/mol, which amounts to about *50%* of the macroscopic susceptibility at room *T*. The spin susceptibility decreases with temperature, by about *25%* between *T=300*K and $T_N$. This deviation to a pure Pauli-constant behavior could be due to the strong correlations present in the 2D Co layer or any *T*-dependence of the Fermi level density of states.

Co NMR can then be used to measure independently the contribution of the two crystallographic Co sites to this susceptibility. We indeed identified two sets of Co NMR signals, called A and B hereafter, with similar weights corresponding to the two Co sites of the unit cell. Their quadrupolar frequencies $v_Q^c = 2.8$ and *4* MHz agree with the findings of Ref. [9]. Site A falls into two subsets $A_1$ and $A_2$ with slightly different $v_Q^c$ and same intensities [10]. The different Co shifts $K_C$ are deduced from the measurement of the different quadrupolar transitions at each temperature. Their T-dependence, plotted on Fig.2, mimics that of $\chi_m$ like Na shifts do. The scaling between $K_C$ and $\chi_m$ leads to $A_{hf}^c = 18 \pm 3 T$ for both A and B sites, and $K_c^{orb} = 1.32; 1.36; 1.57 \pm 0.2\%$ for $A_1$, $A_2$ and B. *The fact that both A and B shifts behave similarly with the same hyperfine coupling demonstrates that neither one can be $Co^{3+}$ above $T_N$.* If Co charge was *3+*, one would expect no *T*-dependent spin shift at all and $K_c^{orb} = 1.9\%$ [1]. Furthermore, the hyperfine coupling $A_{hf}^c$ lies between the values found in $Na_{0.7}CoO_2$ for $Co^{3.3+}$ ( $A_{hf}^c = 4.2\ T$) and $Co^{3.7+}$ ( $A_{hf}^c = 24\ T$), suggesting that the charge of both A and B lies between *3.5±ε* with *ε<0.2* [3]. Similar *T*-dependences for shift and hyperfine couplings were found for $H_0 \perp c$ for site A, confirming our conclusions, while site B was too broad to be resolved in the perpendicular distribution pattern.

Let us now focus on the magnetically ordered regime below $T_N$. Frozen moments develop and induce magnetic splittings of the Co and Na NMR lines associated with the magnetic order. The $A_1$ Co line splits into two subsets shifted by *–h* and *+h* (the B Co signal is not detected anymore, possibly due to a shortening of its transverse relaxation time and/or a large broadening.). Splittings with similar *T*-dependence are observed as well for one Na site, as plotted on fig.3. At the other site, the transferred Co moments cancel with each other, so that it is not splitted.

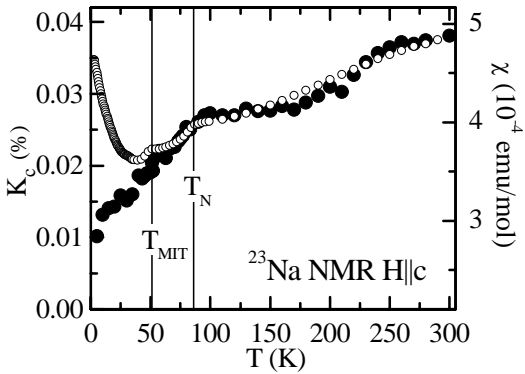

*fig.1 : Na NMR spin shift $K_C$ of both Na sites (left axis, full circles) compared to the macroscopic susceptibility in non oriented powder (right axis, empty circles, H=0.1T).*

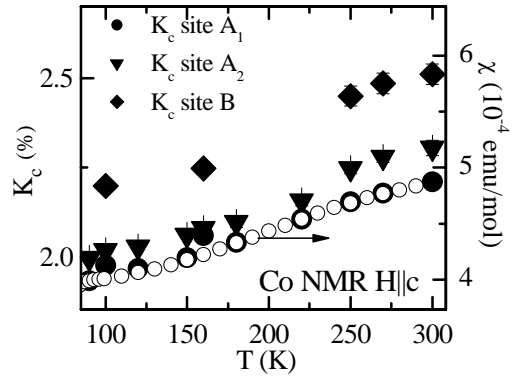

*fig.2 : Co NMR spin shifts $K_C$ for the three Co sites ($A_1$ and $A_2$ correspond to one crystallographic site, and B to the other) (left axis, full symbols) compared to macroscopic susceptibility in non oriented powder (empty circles, right axis).*

In addition to these splittings and contrary to the situation encountered in most antiferromagnets, a paramagnetic component in the magnetization can be detected here, which is found similar for both Na sites. This Na shift measured by the center of gravity of the spectra and plotted in fig.1 is not proportionnal to the macroscopic susceptibility anymore, neither to the frozen moment T-dependence displayed in fig.3.

As Na is coupled to different Co sites from the two Co adjacent layers, the actual moment organization is hard to deduce. However, one can measure the effect of magnetic and MIT transitions on the charge of the Co site through its quadrupolar parameter and on the Co moments through their magnetic splittings. On the charge side, the quadrupolar parameter $\nu_Q^c$ of Co $A_1$ site displayed in fig.3 stays constant below $T=100$K within an incertitude of 0.7%. On the contrary, any CO at $T_N$ or $T_{MIT}$ would result in a change of the EFG on this Co site. In an ionic point charge model, a CO modifying Co layers into alternated chains of $Co^{3+}$ and $Co^{4+}$ should change $\nu_Q^c$ by *19%*. *The constant behavior of $\nu_Q^c$ evidences the absence of any total CO below $T_N$ while magnetic order develops.* Anyhow, a small disproportionnation into $Co^{3.5-\varepsilon}$-$Co^{3.5+\varepsilon}$ could exist, with $\varepsilon<0.2$.

On the magnetic side, the ordering appears commensurate with the lattice. Otherwise, a continuous frequency distribution would be observed instead of the narrow α and β lines for Na NMR. Each Na ion is equally coupled to its two adjacent Co layers. One can thus rule out any simple antiferromagnetic order between planes, which would induce no splitting of all Na lines. The accurate measurement of the Na splitting α done on fig.3 shows a sharp magnetic transition at $T_N$, with a very small change at $T_{MIT}$.

All our results are inconsistent with a charge-ordering or a Mott-like MIT scenario. Not only $T_{MIT}$ and $T_N$ do not coincide, but both results in the metallic and ordered state are incompatible with a simple CO with alternated rows of $Co^{3+}$ and $Co^{4+}$. We propose to explain both the MIT and magnetic transitions by a new scenario, invoking two successive nestings of the Fermi Surface (FS). When two portions of a FS can be nested with each other by a vector $\vec{Q}$, this may lead at low enough temperature to either a spin or charge density wave (SDW, CDW) with a modulation of charge or spin of frequency $\vec{Q}$ [11]. Angle Resolved Photoemission (ARPES) measurements for *x=0.48* reveal a FS with hexagonal shape, as shown in Fig.4 [12]. At *x=0.5*, the specific ordering of Na ions transforms the unit cell from hexagonal to orthorhombic. This in turn changes the Brillouin Zone (BZ) into a smaller rectangle which now cuts the FS as shown in fig.4. In the presence of any small lattice potential, gaps should open at the intersections between BZ and FS and lead to a curvature of the FS as plotted on fig.4. Three different nestings $\vec{Q}_1$, $\vec{Q}_2$ and $\vec{Q}_3$ are now possible. As $\vec{Q}_1$ nests larger regions of the FS, it should occur at higher temperature, i.e. $T_N=86$K. It would naturally lead to a SDW compatible with our present NMR results. As $\vec{Q}_1 \approx (2/3; 1/3)$ in hexagonal reciprocal cell units, it is commensurate with the Co lattice and must lead to a commensurate SDW, as observed here. The magnetic moment *T*-dependence in fig.3 shows an increase near $T_C$ slightly sharper than in the weak-coupling BCS limit, similar to other SDW systems [11]. The associated opening of a gap on the nested parts of the FS is observed in optics below *T~100*K [5] as well as in resistivity [4].

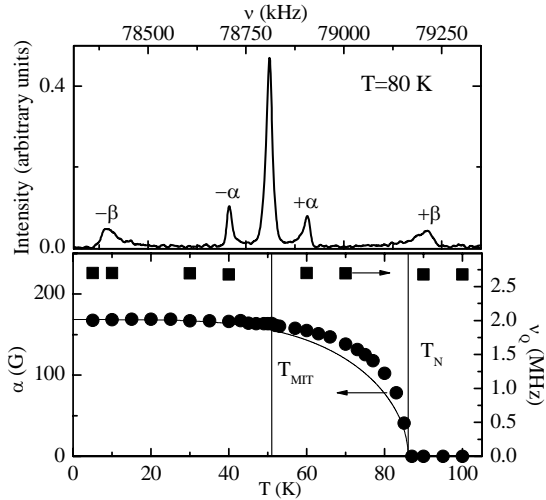

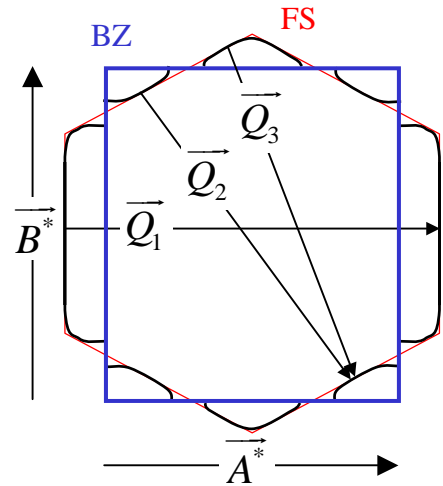

Fig.3 : *Upper panel* : Na NMR central transition below $T_N$ where frozen moments induce a ± α and ± β splitting of part of the central line. *Lower panel, left axis* : the T-dependence of the splitting α, which is proportional to the *c*-component of the local field resulting from frozen moments (circles) together with a BCS weak-coupling fit (line). *Right axis* : the quadrupolar frequency $\nu_Q^c$ of Co A site proportional to electric field gradient along *c* axis.

Fig.4 : Extended zone scheme of the reciprocal lattice: at *x=0.5*, the Brillouin Zone (BZ) is a rectangle (blue contour) due to the Na ordering, with reciprocal unit vectors A* and B*. This should transform the hexagonal Fermi Surface FS measured in Ref. [11] at *x=0.48* (red) into the black contour. In this new FS, three possible nestings $\vec{Q}_1$, $\vec{Q}_2$ and $\vec{Q}_3$ are identified.

The carriers responsible for transport are now linked to the non-nested corners of the FS, which explains then the change of sign of the Hall coefficient at $T_N$. The next transition at $T_{MIT}=51$K originates from a second nesting between the residual pockets. This nesting is expected to occur at $\vec{Q}_3$ more likely than $\vec{Q}_2$ since it nests larger regions of FS. But ARPES measurements are needed to determine the exact nesting vector which will be very dependent on the actual shape of the pockets. This second nesting would leave very few carriers at Fermi Level, explaining the sharp "MIT" increase in resistivity and the observed large vanishing of the FS [13]. The presence of residual small pockets would then be due to the non-nested parts of FS, which could be partly responsible for the observed susceptibility measured by Na NMR below $T_N$. If this nesting leads to a SDW as well, the resulting magnetic order would then mix-up both $\vec{Q}_1$ and $\vec{Q}_3$ modulations, likely with a much smaller weight associated to $\vec{Q}_3$. Hence only small modification to the existing moments would be observed at $T_{MIT}$, in agreement with our findings, while magnetic ordering could be slightly affected, as observed by muon spin resonance ($\mu SR$) [6]. The anomalies detected also at $T=30$K both in resistivity and magnetism [2,6] could be linked to a further instability here again linked to a nesting of the remaining pockets. Electron diffraction studies reveal the presence of an additional superstructure observed below $T=100$K [2]. This may stem from a coupling between the two successive SDW and the lattice, similarly to that observed in Cr or organic compounds [14]. The succession of two CDW driven by two nestings has been observed in 1D or 2D metals such as $NbSe_3$ or monophosphate tungsten bronzes, together with a change in sign of the Hall coefficient very similar to the present case [15]. To our knowledge, $Na_{0.5}CoO_2$ would be the first case where SDW are involved.

In conclusion, our NMR results allow us to evidence that the $Co^{3+}$-$Co^{4+}$ charge order scenario does not apply in the Na-ordered $Na_{0.5}CoO_2$. We assign the magnetic transition to a SDW associated with a nesting of the FS. This explains both our NMR results and previous Hall, resistivity and optical conductivity measurements. We attribute the MIT transition to a second nesting of the remaining FS pockets which would explain the weak modifications of the magnetic order, while FS and resistivity are strongly affected. In our view, the $x=0.5$ composition is peculiar in the phase diagram because the corresponding Na-ordering favors specific nestings of the FS. In this scenario, the large Coulomb interaction U present in Co layers naturally provides the coupling necessary for the appearance of a SDW, and $Na_{0.5}CoO_2$ definitely belongs to the class of low dimensional strongly correlated materials. The magnetic orders observed for $x>0.75$ with small magnetic moments could be explained in a SDW scenario as well. Here again, the associated Na orderings could trigger different nesting vectors, and explain why $T_N$ does not evolve monotonously with Na content. ARPES measurements are clearly needed to establish how the FS evolves with temperature and Na doping.

*Note added*: During completion of the manuscript, we got aware of a Co NMR and neutron study on the same material focussed mostly on the magnetic state, which corroborates our results [16]. However, in the proposed magnetic structure, no Na splitting should be observed in contrary to our measurements. Another NMR study [17] submitted just after ours concludes that the 23Na NMR data also supports the absence of charge ordering at the magnetic transition. However, the sensitivity of the Na EFG to Co charge and the accuracy of their data are not sufficient to allow the authors to reach such conclusions independently of our results. An ARPES study submitted after ours reports the detection of a reconstruction of the Fermi Surface due to Na ordering which appears consistent with our nesting scenario [18].

*We acknowledge V. Brouet, N. Dupuis, P. Foury, M. Heritier, D. Jerome, P. Lederer, P. Mendels, I. Mukhamedshin, D. Nunez-Regueiro, J.P. Pouget, S. Ravy and P. Wzietek for fruitful discussions.*